\documentclass[%
groupedaddress,
preprint,
 amsmath,amssymb,
 aps,
prl
]{revtex4-1}
\usepackage{graphicx}
\usepackage{dcolumn}
\usepackage{bm}

\newcommand{\phext}{\ensuremath{\varphi_\mathrm{ext}}}
\newcommand{\taua}{\ensuremath{\tau_\mathrm{avg}}}
\newcommand{\hebt}{\ensuremath{H_{\mathrm{EB}}(\varphi_{\mathrm{ext}})}}
\newcommand{\hct}{\ensuremath{H_\mathrm{C}(\varphi_{\mathrm{ext}})}}
\newcommand{\heb}{\ensuremath{H_\mathrm{EB}}}
\newcommand{\hib}{\ensuremath{\vec{H}_\mathrm{IB}}}
\newcommand{\hfc}{\ensuremath{\vec{H}_\mathrm{FC}}}
\newcommand{\hc}{\ensuremath{H_\mathrm{C}}}
\newcommand{\kf}{\ensuremath{K_\mathrm{F}}}
\newcommand{\tf}{\ensuremath{t_\mathrm{F}}}
\newcommand{\irmn}{Ir$_{17}$Mn$_{83}$}
\newcommand{\cofe}{Co$_{70}$Fe$_{30}$}
\newcommand{\betf}{\ensuremath{\beta_{\mathrm{F}}}}

\newcommand{\aleb}{\ensuremath{\gamma_{\mathrm{EB}}}}
\newcommand{\alra}{\ensuremath{\gamma_{\mathrm{RMA}}}}
\newcommand{\alf}{\ensuremath{\gamma_{\mathrm{F}}}}

\newcommand{\msat}{\ensuremath{M_{\mathrm{sat}}}}
\newcommand{\hext}{\ensuremath{H_{\mathrm{ext}}}}
\newcommand{\vhext}{\ensuremath{\vec{H}_{\mathrm{ext}}}}
\newcommand{\gra}{\ensuremath{^{\circ}}}

\newcommand{\jeb}{\ensuremath{J_\mathrm{EB}^\mathrm{eff}}}
\newcommand{\jhc}{\ensuremath{J_\mathrm{C}^\mathrm{eff}}}
\newcommand{\pib}{$\uparrow \uparrow$IB}
\newcommand{\aib}{$\uparrow \downarrow$IB}

\begin{document}


\title{Quantitative analysis of the influence of keV He ion bombardment on exchange bias layer systems}

\author{Nicolas David M{\"u}glich}
\email[]{nicolas.mueglich@gmail.com}
\affiliation{Institute of Physics and Center for Interdisciplinary Nanostructure Science and Technology (CINSaT),
University of Kassel, Heinrich-Plett-Strasse 40, 34132 Kassel, Germany}

\author{Gerhard G{\"o}tz}

\affiliation{Center for Spinelectronic Materials and Devices, Physics Department, Bielefeld University, Universit{\"a}tsstra{\ss}e 25, 33615 Bielefeld, Germany}

\author{Alexander Gaul}
\author{Markus Meyl}

\affiliation{Institute of Physics and Center for Interdisciplinary Nanostructure Science and Technology (CINSaT),
University of Kassel, Heinrich-Plett-Strasse 40, 34132 Kassel, Germany}


\author{G{\"u}nter Reiss}

\author{Timo Kuschel}

\altaffiliation[Present address: ]{Physics of Nanodevices, Zernike Institute for Advanced Materials, University of Groningen, Nijenborg 4, 9747 Groningen, The Netherlands}

\affiliation{Center for Spinelectronic Materials and Devices, Physics Department, Bielefeld University, Universit{\"a}tsstra{\ss}e 25, 33615 Bielefeld, Germany}
\author{Arno Ehresmann}
\affiliation{Institute of Physics and Center for Interdisciplinary Nanostructure Science and Technology (CINSaT),
University of Kassel, Heinrich-Plett-Strasse 40, 34132 Kassel, Germany}

\date{\today}

\begin{abstract}
The mechanism of ion bombardment induced magnetic patterning of exchange bias layer systems for creating engineered magnetic stray field landscapes is still unclear. We compare results from vectorial magneto-optic Kerr effect measurements to a recently proposed model with time dependent rotatable magnetic anisotropy. Results show massive reduction of rotational magnetic anisotropy compared to all other magnetic anisotropies. We disprove the assumption of comparable weakening of all magnetic anisotropies and show that ion bombardment mainly influences smaller grains in the antiferromagnet.
\end{abstract}


\maketitle


\section{Introduction} 
Polycrystalline double layers of an antiferromagnet (AF) and a ferromagnet (F) show a unidirectional magnetic anisotropy (UDA), which is called exchange bias (EB) \cite{Mei56_mod,mei57_new}. The resulting pinned magnetization direction is commonly used in magnetic sensors as a reference electrode \cite{Nog99_rev}. With the possibility to modify the local direction of the UDA by light ion bombardment induced magnetic patterning (IBMP \cite{Mou01_ib,Mou01_struk,juraszek2002tuning,Ehr04_pattern,Ehr06_pattern}), micromagnetic configurations with designed magnetic stray field landscapes can be tailored \cite{holzinger2013tailored,gaul2016engineered}. This opens a door to a whole set of new applications, which, e.g., will be a central part of magnetic particle transport in lab-on-chip applications for biosensing \cite{Ehr11_asym,Hol15_dir,Hol15_mix,Ehr15_sensor}.
\\Among various models trying to explain EB \cite{malozemoff1987random,mauri1987simple,stamps2000mechanisms}, one class of models is particularly suited to describe EB in polycrystalline layer systems using a classification of AF grains into different categories of thermal stability \cite{soeya1996nio,Ehr05_mod,OGr10_york}. In the Stoner-Wohlfarth-approach \cite{Sto47_mod,Sto48_mod} on the other hand, the magnetic anisotropies induced by thermally stable and unstable grains are modeled with UDA and rotational magnetic anisotropy (RMA), respectively \cite{Ges02_rot,Har12_model}. Recently, we could improve these coherent rotation models introducing the average relaxation time of thermally unstable grains as a parameter to describe the contribution of RMA \cite{Mug16_rma}. Calculations of the EB field \heb~and coercive field \hc~using this approach have been compared to experiments on magnetization reversal as a function of external magnetic in-plane field angle. The new model described the experiments very well and allowed a meaningful determination of material parameters. The model may enable additionally to investigate the microscopic mechanisms of IBMP when determining the change of material parameters upon light-ion bombardment (IB).
\\IBMP is caused by hyperthermal energy transfer from He ions to the EB system, layer intermixing and defect creation \cite{Ehr05_mod,Ehr11_drift}. These sources of energy deposition may change all involved anisotropies of the EB layer system, such as the well known modification of the EB itself. The different processes and their possibly different consequences on magnetic characteristics of EB layer systems, however, have as yet not been disentangled because it is not possible to experimentally investigate them individually. Nevertheless, with detailed knowledge of IB it might be possible to influence the different anisotropies individually leading to an improved usability of EB systems, e.g. reducing coercivity without changing the magnitude of \heb.
\\In this work, we investigate the influence of IB on the magnetic properties of EB layer systems quantitatively by analysing the magnetization measured under variation of the in-plane external magnetic field angle \phext~for different He ion doses $D$. By fitting the magnetic material properties of the model given in Ref. \cite{Mug16_rma} to the experimental data, i.e. the relations \hebt~and \hct, a dose-dependent relation of the magnetic material properties becomes available. We show that IB does not lead to a simultaneous and comparable weakening of all magnetic anisotropies; instead it is shown that IB mainly influences smaller grains in the AF.

\section{Experimental}
EB samples of the type Cu$^{50~\mathrm{nm}}$/\irmn$^{30~\mathrm{nm}}$/\cofe$^{15~\mathrm{nm}}$/Si$^{20~\mathrm{nm}}$ were fabricated on naturally oxidized Si including a field cooling procedure with a setting temperature of 573.15~K and an external magnetic field \hfc~of 80~kA/m as described in Ref. \cite{Mug16_rma}. Thereafter, samples were bombarded with 10 keV He ions in a home built setup described in Ref. \cite{lengemann2012plasma} under high vacuum conditions with a base pressure of $4\cdot10^{-6}$~mbar. IB was carried out with an ion current of approximately $10^{-6}$~mA and $D$ ranging from $10^{13}$ to $10^{16}$~ions/cm$^2$. During IB an external magnetic field \hib~of 70~kA/m was applied either parallel (\pib) or antiparallel (\aib) to \hfc. 
\\Magnetization curves as a function of \phext~were measured using vectorial magneto-optic Kerr effect (MOKE) magnetometry as described in Ref. \cite{kuschel2011vectorial,jimenez2014vectorial,Mug16_rma}. In the setup longitudinal and transversal components of the magnetization are detected by analyzing polarization and intensity of the reflected light from a 632~nm diode laser, respectively. Magnetization curves were measured using 300 external magnetic field steps in a range of $\pm$80~kA/m within 80 seconds. For each sample \phext~was varied in the range of 450\gra~with a resolution of 1\gra. Absence of significant training effects was verified by comparing data from 0\gra~to 90\gra~with 360\gra~to 450\gra.

\section{Model}
\begin{figure}[htbp]
\includegraphics{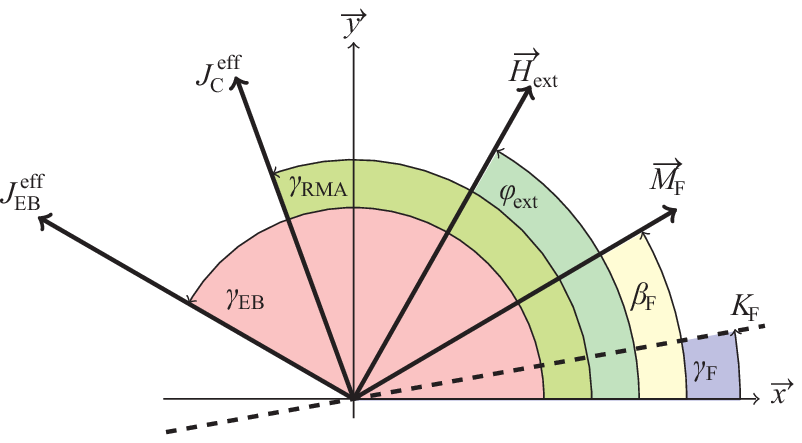}
\caption{\label{modelpic2} Graphical illustration of angles and vectors for the used model in a polar coordinate system. $\vec{M}_\mathrm{F}$~is the magnetization vector of the F layer and \betf~its azimuth. \kf~is the energy density per unit volume of the FUMA and \alf~the azimuth of its magnetic easy direction. \vhext~is the external magnetic field vector with the azimuth \phext. \jeb~(\jhc) is the energy area density of the UDA (RMA) with the corresponding azimuth \aleb~(\alra). Taken from Ref. \cite{Mug16_rma}.}
\end{figure}

For the numerical calculations the model of Ref. \cite{Mug16_rma} was used, which is based on the coherent rotation approach from Stoner and Wohlfarth \cite{Sto47_mod,Sto48_mod} calculating the angle of the F magnetization direction \betf~ with respect to the $x$-axis (see FIG. \ref{modelpic2} for an overview of magnetic anisotropy and angle definitions). There, magnetic anisotropy of the F was assumed to be uniaxial (FUMA) with the energy volume density \kf~while the influence of the AF was modeled with two different magnetic anisotropy energies connected to different classes of grains \cite{soeya1996nio,Ehr05_mod,OGr10_york} in the AF. While grains of Class I and IV do not induce anisotropy in the coherent rotation model, influence of thermally stable grains of Class III is taken into account by UDA. Influence of thermally unstable grains (Class II) is modeled with RMA by considering an average relaxation time \taua~for all grains of the corresponding class. The energy area density of UDA and RMA is \jeb~and \jhc, respectively, and is connected to the number of AF grains in each category. The energy area density of the system in total is
\begin{equation}
\label{eq_full}
\begin{aligned}
E/A=&(E_\mathrm{Z}+E_\mathrm{FUMA}+E_\mathrm{UDA}+E_\mathrm{RMA})/A\\
=&-\mu_0\hext\msat\tf\cos{\left(\betf-\phext\right)}\\
&+\kf\tf \sin^2{\left(\betf-\alf\right)}\\
&-\jeb \cos{\left(\betf-\aleb\right)}\\
&-\jhc \cos{\left(\betf-\alra\right)}\\
\mathrm{with}~ \alra &= \betf(t-\taua).
\end{aligned}
\end{equation}
Here, $\mu_0$ is the magnetic permeability in vacuum, \msat~the saturation magnetization of the F with thickness \tf. \hext~and \phext~are strength and direction of the external magnetic field, respectively. \alf, \aleb~and \alra~are the angles between the $x$-axis of the coordinate system and the magnetic easy directions of FUMA, UDA and RMA, respectively (cf Fig \ref{modelpic2}). While \alf~and \aleb~emerge from sample processing \alra~is connected to \betf~via \taua~with the time $t$ \cite{Mug16_rma}.

\section{Results}
\begin{figure}[htbp]
\includegraphics{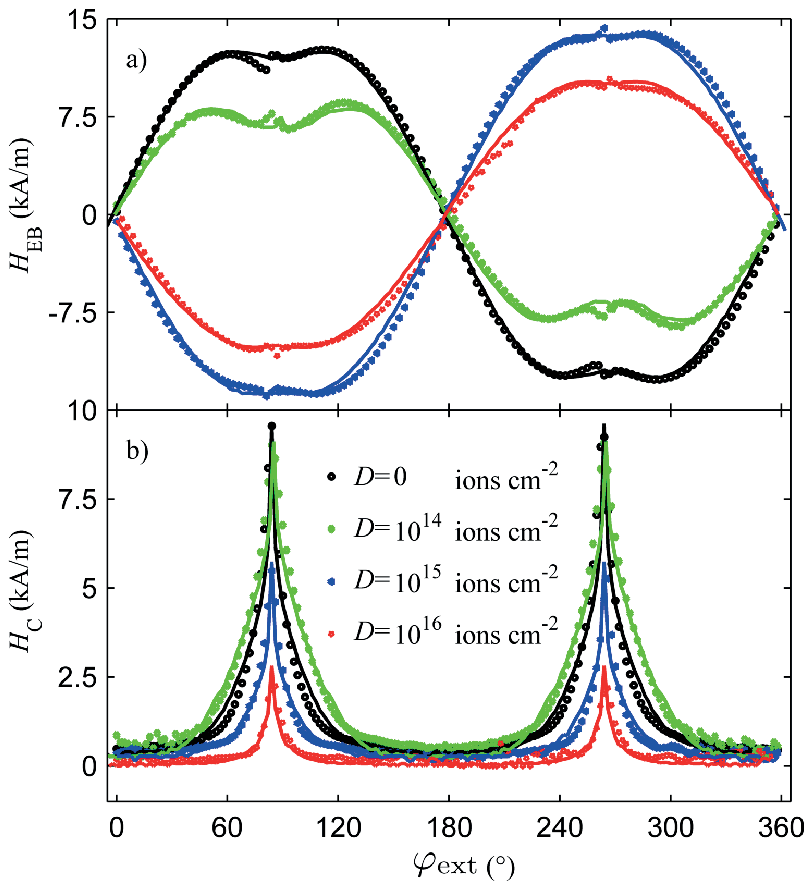}
\caption{\label{iondose_anti}Experimentally determined (a) \hebt~and (b) \hct~for an EB system of Cu$^{50~\mathrm{nm}}$/\irmn$^{30~\mathrm{nm}}$/\cofe$^{15~\mathrm{nm}}$/Si$^{20~\mathrm{nm}}$ after \aib~with different $D$. Solid lines correspond to a fit using Eq. \ref{eq_full}.}
\end{figure}

\begin{figure}[!htbp]
\includegraphics{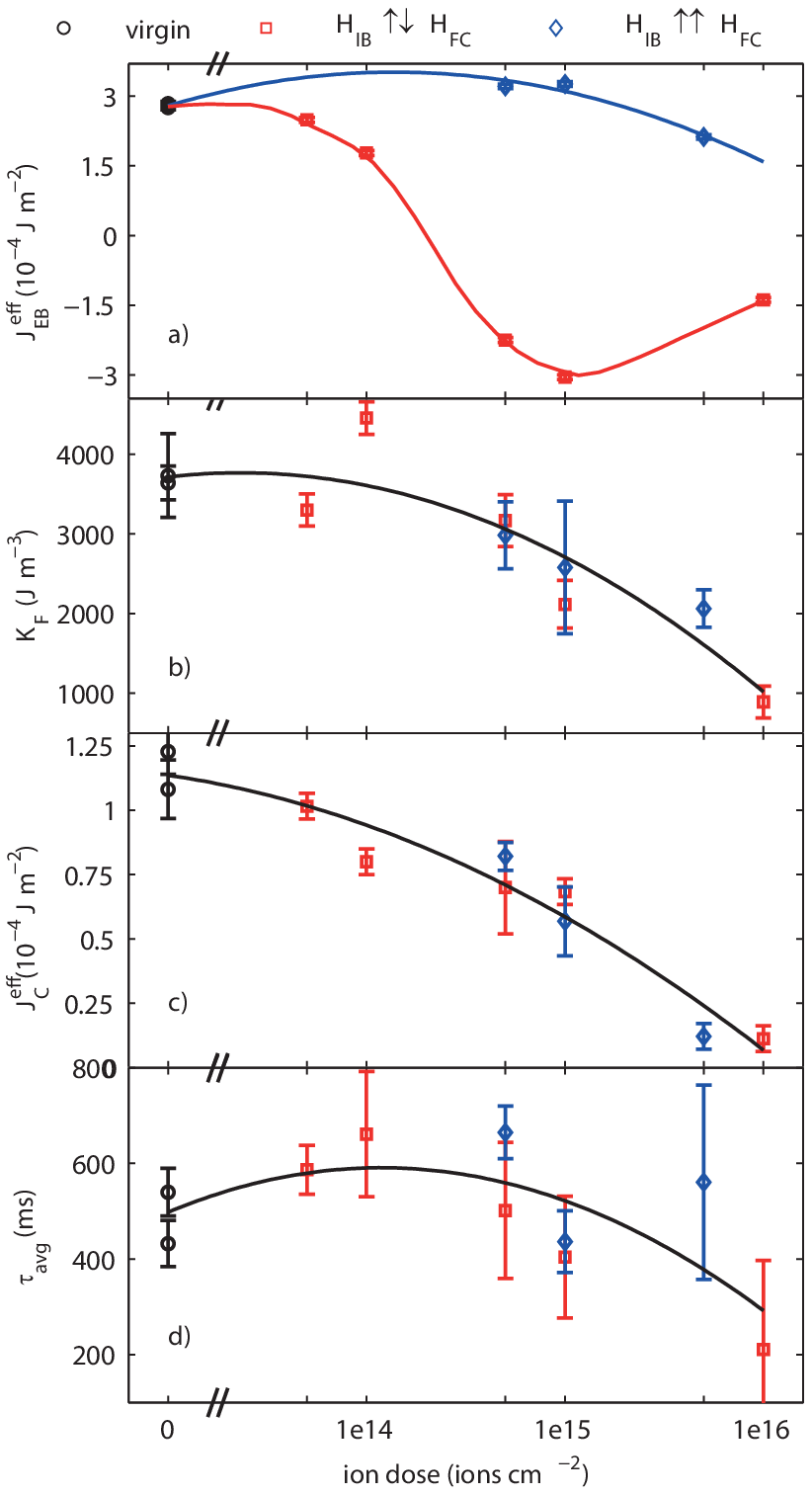}
\caption{\label{ionpara_anti} Characteristic quantities (a) \jeb, (b) \kf, (c) \jhc~and (d) \taua~of the EB layer in dependence on $D$ obtained by a fit of the model parameters to the experimental relations \hebt~and \hct. Black circles belong to unbombarded, red squares and blue diamonds to bombarded samples in \aib~and \pib~configuration, respectively. The error corresponds to the uncertainty in the fit constants introduced by a variation of the starting conditions by a factor of 3. Lines are a guide to the eye.}
\end{figure}

\begin{figure}[!htbp]
\includegraphics{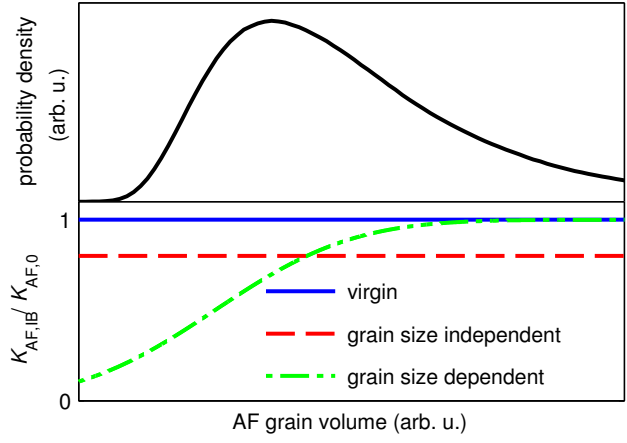}
\caption{\label{prob_function}(a) Sketch of the grain size distribution in the AF layer using a lognormal function. (b) Sketch of the relative change of the magnetic anisotropy due to IB $K_\mathrm{AF,IB}/K_\mathrm{AF,0}$ in dependence on AF grain volume for a virgin sample (blue line), grain size independent (red/dashed) and dependent (green/dash-dotted) reduction of $K_\mathrm{AF}$.}
\end{figure}

\begin{figure}[!htbp]
\includegraphics{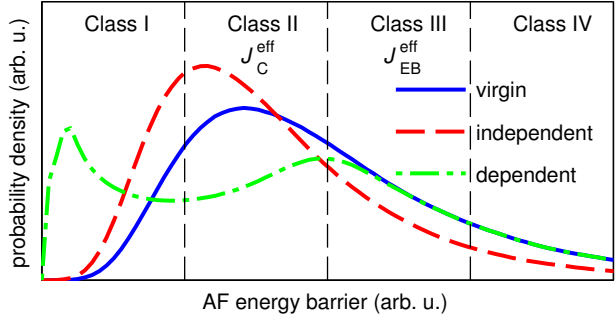}
\caption{\label{energie_barriers}Sketch of the energy barrier probability density in dependence on AF grain volume for a virgin sample (blue line), grain size independent (red/dashed) and dependent (green/dash-dotted) reduction of $K_\mathrm{AF}$.}
\end{figure}

The relations \hebt~and \hct~determined from vectorial MOKE measurements for different $D$ (see FIG. \ref{iondose_anti}) show the modifications of the characteristic quantities \heb~and \hc~of the investigated EB layer systems by IB. \hebt~behaves roughly sinusoidal for all ion doses. The fine structure near the magnetic easy axis of the system ($\phext\approx90\gra$) leading to a deviation from the sine shape is caused by FUMA \cite{xi1999study,Bai10_angle,Mug16_rma}. Small deviations from the mirror symmetry with respect to the magnetic easy axis can be attributed to a small misalignment ($\leq2\gra$) between FUMA and UDA \cite{Jim09_noncoll,Jim11_aniso}. For increasing $D$ in the \aib~case the amplitude of \hebt~is successively reduced ($D=10^{14}~\mathrm{ions~cm}^{-2}$) until it reaches a maximum with opposite sign ($D=10^{15}~\mathrm{ions~cm}^{-2}$) and is subsequently reduced in magnitude ($D=10^{16}~\mathrm{ions~cm}^{-2}$). This data is in accordance with previous magnetic easy axis hysteresis measurements \cite{Mou01_ib,juraszek2002tuning,Eng03_aniso}. In addition, the fine structure near the magnetic easy axis of the system is reduced for higher ion doses and almost vanishes for the highest ion dose so \hebt~almost forms a sine function indicating a reduction of the FUMA.
\\The experimentally determined relations \hct~show a lorentzian like shape indicating that coercivity results from a superposition of FUMA and RMA \cite{Mug16_rma}. Peak height and width of \hct~is reduced in general with higher $D$.
\\A fit of the model parameters to the experimental data (see solid lines in FIG. \ref{iondose_anti}) reveals material parameters as functions of $D$ (see FIG. \ref{ionpara_anti}). For the calculations \msat~was assumed to be dependent on the ion dose with 
\begin{equation}
\msat=\left(1230-5\cdot10^{-14}\cdot D\cdot\mathrm{\frac{cm^{2}}{ions}}\right)~\mathrm{kA/m}
\end{equation} as derived in Ref. \cite{huckfeldt2017modification}. \jeb~as a function of $D$ is the only relation showing a dependence on the direction of \hib (cf Fig. \ref{ionpara_anti} (a)). In the \pib~case \jeb~is slightly increased for smaller $D$ while it is more complex in the \aib~case. Both relations are in accordance to previous studies where \heb~was measured by magnetic easy axis hysteresis loops as a function of $D$ \cite{Mou01_ib,Ehr11_drift}. The small increase in the absolute value of \jeb~at $D\approx10^{15}~\mathrm{ions~cm}^{-2}$ is linked to a part of Class IV grains (FIG. \ref{energie_barriers}). Due to the high  energy transfer of the ions into the layer system some of these grains may overcome their relatively high energy barrier and relax into their energetically favored state. The decrease of \jeb~for higher ion doses of $D\approx10^{16}~\mathrm{ions~cm}^{-2}$ is linked to a general decrease of the F/AF interaction as a consequence of the interlayer intermixing becoming dominant for high ion doses \cite{schmalhorst2004x}. The complex structure of \jeb~in the \aib~case including the sign change is an effect of the local field cooling induced by the hyperthermal energy transfer of the He ions into the layer system. One should note that the negative sign of \jeb~refers to the average macroscopic \jeb~direction determined relatively to the average \jeb~of the unbombarded sample. The sign of the microscopic \jeb~is not changed. In contrast to previous magnetic easy axis hysteresis measurements the angular resolved measurements are clearly showing that EB reorientation takes place successively on a local basis. Reorientation of the UDA does not take place via coherent rotation. 
\\\kf~as the characteristic quantity of FUMA shows a decrease for higher $D$ and is reduced to roughly 25~\% of the initial value for $D=10^{16}~\mathrm{ions~cm}^{-2}$ (cf Fig. \ref{ionpara_anti} (b)) which is in qualitative accordance to previous studies of a different material system of a pure F \cite{chappert1998planar}. The reduction of anisotropy can be attributed to the ion induced defect creation in the F layer leading to a decrease of crystalline order. \kf~shows no dependence on the direction of \hib.
\\For \jhc~and \taua, which are used to describe the thermally unstable grains of Class II, again, no influence of the \hib~direction on the dose dependency was detected (cf Fig. \ref{ionpara_anti} (c), (d))). This was suspected although the magnetic state of these grains is affected by the external field during IB because the magnetic state is thermally unstable and, therefore, the magnetic conditions during the field cooling process are not memorized. \jhc~shows a massive decrease to only 10~\% of the initial value for higher ion doses of $D=10^{16}~\mathrm{ions~cm}^{-2}$. For small ion doses \taua~is increased slightly while it is not possible to determine a trend for higher ion doses due to the large uncertainty. As the influence of RMA decreases rapidly with higher ion dose a precise determination of its time dependency is not possible anymore.
\\Summing up IB suppresses RMA much more than UDA (by factor of 5 at $D=10^{16}~\mathrm{ions~cm}^{-2}$). This is a surprising result if one would assume a statistical distribution of defects in all AF grains as we will see in the following discussion. In first order, the energy densities of both, RMA and UDA, can be approximated as the product of two quantities. Firstly, the number of grains in the respective Class (II for RMA, III for UDA) and secondly, the microscopic F/AF exchange interaction constant $J_{\mathrm{F/AF,i}}$. Both are altered by defect creation as IB not only decreases $J_{\mathrm{F/AF,i}}$ but also the magnetic anisotropy of AF grains $K_\mathrm{AF}$ \cite{Ehr05_mod}. Thus, thermal stability of AF grains is changed leading to different numbers of grains in each category. 
\\Let us discuss how UDA and RMA would evolve in case of evenly distributed defects. Here, $J_{\mathrm{F/AF,i}}$ and $K_\mathrm{AF}$ should be reduced by a similar percentage for all grain sizes as the number of defects per unit volume is constant. Consequently, different behavior of UDA and RMA could only be related to the number of grains in the respective categories. For a lognormal AF grain size distribution (see FIG. \ref{prob_function}(a)) the resulting distribution of AF energy barriers should be also lognormal (blue line  in FIG. \ref{energie_barriers}). Grain size independent reduction of $K_\mathrm{AF}$ (dashed line in FIG. \ref{prob_function}(b)) would shift this distribution to lower energies (dashed line in FIG. \ref{energie_barriers}). Obviously such a shifted energy barrier distribution is not able to explain the dramatic decrease of \jhc~compared to \jeb~for higher $D$. This holds even if all grains of Class IV would contribute to \jeb~as the number of unset grains at field cooling temperatures of 300\gra~for an AF with a N\'{e}el temperature of 400\gra~should be rather low \cite{OGr10_york}. 
\\The situation is different if IB would mainly influence smaller AF grains. Grain size dependent reduction of $K_\mathrm{AF}$ (dash-dotted line in FIG. \ref{energie_barriers}) would result in many Class II grains becoming superparamagnetic while bigger grains are nearly unaffected. This would not only explain the massive decrease of \jhc, but also the increase of \taua~for smaller ion doses as the average energy barrier of the remaining Class II grains is increased. We can see two reasons why reduction of $K_\mathrm{AF}$ by IB predominantly  takes place in smaller grains. Firstly, smaller grains could be less thick, i.e. their spatial extension orthogonal to the F/AF interface is smaller, which makes them prone to be affected by intermixing. Secondly, their proportion of surface atoms is higher. At room temperature the chance of recombination after defect creation in bulk is 99~\% \cite{Ziegler2008} so that most of the created defects vanish quickly. In case of surface atoms displaced atoms could be out of range for recombination reducing their chance of recombination.
\\Summing up, the influence of IB cannot be described by a comparable weakening of all magnetic anisotropies. In fact, IB seems to have a stronger effect on smaller grains with an increased proportion of surface atoms and/or a smaller thickness.

\section{Summary}
In this work, we have determined the influence of keV He IB on the material properties of EB systems quantitatively for the first time by fitting a coherent rotation model to experimental data obtained by vectorial MOKE magnetometry. The results show that IB induced reversal of EB takes place locally instead of a coherent rotation underpinning the validity of the polycrystalline model of EB. The reduction of coercivity caused by the IB can be attributed to both a reduction of FUMA and RMA. The fast suppression of RMA with increasing ion doses suggests that the reduction of magnetic anisotropy in the AF is more prominent in smaller grains having a higher proportion of surface atoms and/or a smaller thickness. This excludes the assumption that IB causes a comparable weakening of all magnetic anisotropies. Our work shows the functionality of IB as it allows not only magnetic patterning but also eliminates the coercivity inducing RMA which is disrupting in most applications.
\section{Acknowledgments}
NDM thanks the University of Kassel for the Universit\"at Kassel Promotionsstipendium.
\bibliography{vmoke2}

\end{document}